\documentclass[12pt]{iopart}

\usepackage{iopams}
\usepackage{graphicx}

\usepackage[square,numbers,sort&compress]{natbib}

\usepackage{xcolor}

\definecolor{TangoButter1}{HTML}{FCE94F}
\definecolor{TangoButter2}{HTML}{EDD400}
\definecolor{TangoButter3}{HTML}{C4A000}
\definecolor{TangoOrange1}{HTML}{FCAF3E}
\definecolor{TangoOrange2}{HTML}{F57900}
\definecolor{TangoOrange3}{HTML}{CE5C00}
\definecolor{TangoChocolate1}{HTML}{E9B96E}
\definecolor{TangoChocolate2}{HTML}{C17D11}
\definecolor{TangoChocolate3}{HTML}{8F5902}
\definecolor{TangoChameleon1}{HTML}{8AE234}
\definecolor{TangoChameleon2}{HTML}{73D216}
\definecolor{TangoChameleon3}{HTML}{4E9A06}
\definecolor{TangoSkyBlue1}{HTML}{729FCF}
\definecolor{TangoSkyBlue2}{HTML}{3465A4}
\definecolor{TangoSkyBlue3}{HTML}{204A87}
\definecolor{TangoPlum1}{HTML}{AD7FA8}
\definecolor{TangoPlum2}{HTML}{75507B}
\definecolor{TangoPlum3}{HTML}{5C3566}
\definecolor{TangoScarletRed1}{HTML}{EF2929}
\definecolor{TangoScarletRed2}{HTML}{CC0000}
\definecolor{TangoScarletRed3}{HTML}{A40000}
\definecolor{TangoAluminium1}{HTML}{EEEEEC}
\definecolor{TangoAluminium2}{HTML}{D3D7CF}
\definecolor{TangoAluminium3}{HTML}{BABDB6}
\definecolor{TangoAluminium4}{HTML}{888A85}
\definecolor{TangoAluminium5}{HTML}{555753}
\definecolor{TangoAluminium6}{HTML}{2E3436}
\definecolor{LightGray}{HTML}{CCCCCC}

\newcommand{\hl}[1]{#1}

\newcommand{\est}{rate\ of\ adaptation}
\begin{document}

\title[Convection shapes resistance evolution in spatial gradients]{Convection shapes the trade-off between antibiotic efficacy and the selection for resistance in spatial gradients%
}
 
\author{Matti Gralka$^\dag$, Diana Fusco$^{\dag\ddag}$, Stephen Martis$^\dag$ and Oskar Hallatschek$^{\dag\ddag}$\footnote{To whom correspondence should be addressed.}}

\address{\dag\ Department of Physics, University of California, Berkeley, CA 94720}

\address{\ddag\ Biophysics and Evolutionary Dynamics Group, Departments of Physics and Integrative Biology, University of California, Berkeley, CA 94720}

\begin{abstract}
Since penicillin was discovered about 90 years ago, we have become used to using drugs to eradicate unwanted pathogenic cells. However, using
 drugs to kill bacteria, viruses or cancer cells has the serious side effect of selecting for mutant types that survive the drug attack. A key question therefore is
how one could eradicate as many cells as possible for a given acceptable risk of drug resistance evolution. We address this general question in a model of drug resistance evolution in spatial drug gradients, which recent experiments and theories have suggested as key drivers of drug resistance.  Importantly, our model takes into account the influence of convection, resulting for instance from blood flow. Using stochastic simulations, we study the fates of individual resistance mutations and quantify the trade-off between the killing of wild-type cells and the rise of resistance mutations: shallow gradients and convection into the antibiotic region promote wild-type death, at the cost of increasing the establishment probability of resistance mutations. 
We can explain these observed trends by modeling the adaptation process as a branching random walk. Our analysis reveals that the trade-off between death and adaptation depends on the relative length scales of the spatial drug gradient and random dispersal, and the strength of convection. Our results show that convection can have a momentous effect on the rate of establishment of new mutations, and may heavily impact the efficiency of antibiotic treatment.
\end{abstract}
\newpage
\section{Introduction}
The emergence of drug resistance represents one of the major clinical challenges of the current century~\cite{Neu1992,Spellberg2008,davies2010origins}. Microbial pathogens quickly acquire resistance to new antibiotics~\cite{Levy2004}, while solid tumors often regrow after treatment because of resistance mutations that arise during tumor growth~\cite{Lambert2011}. In addition to genomic studies examining the molecular causes of resistance~\cite{raguz2008resistance,Blair2015}, the dynamics of drug resistance evolution has recently attracted wide interest~\cite{Toprak2012,allen2016antibiotic}, with the dual goal of understanding the emergence of resistance and developing novel strategies to prevent or control its spread~\cite{Ramsayer2013,maclean2010population}. Next-generation sequencing and high-throughput experimental techniques enable the quantitative study of resistance evolution but require the development of new theories to appropriately interpret experimental results~\cite{Baym2016}. 

In many realistic systems, an evolving population interacts with its surroundings and exhibits a well-defined spatial structure (for instance, in tumors and biofilms~\cite{Lambert2011}). It has recently been shown that this spatial structure can strongly influence the subclonal structure and the adaptation of spatially expanding populations, both from \textit{de novo} and pre-existing mutations~\cite{Ling2016,Fusco2016,Gralka2016}. In addition, the presence of spatial drug gradients is well documented both in the outside environment~\cite{Martinez2009,Andersson2014} as well as within biofilms~\cite{Stewart2001} and the human body~\cite{Nix1991,Elliott1995,meulemans1989measurement,minchinton2006drug}. The presence of spatial~\cite{kepler1998drug,debarre2009evolutionary,fu2015spatial,Zhang1764,Bell2011a,Baym2016} and temporal~\cite{Perron2008,Lindsey2013,Ramsayer2013,Bell2009} heterogeneities has been shown to facilitate the emergence of drug-resistant phenotypes and enable populations to reach a higher degree of resistance than in homogeneous drug concentrations. For instance, in a microfluidic experiment, a spatial gradient gave rise to a higher rate of adaptation of bacterial populations~\cite{Zhang1764}. Similarly, microbes growing on soft agar plates with gradually increasing antibiotic concentrations were able to rapidly evolve resistance to high levels of antibiotics, while sudden jumps to unsustainably high concentrations dramatically slowed down adaptation~\cite{Baym2016}. 
Moreover, many realistic growth scenarios of bacterial populations may be subject to convection driving them up or down the gradient. Examples include the gut, arteries, and urethra in the human body~\cite{Passerini1992,Hall-Stoodley2004,Persat2015,Cremer2016}, but also flows in aquatic environments, like ocean and river currents~\cite{Battin2007}, or flow in pipes and catheters~\cite{Stickler2008}. 

A number of recent theoretical studies have investigated how gradients speed up the evolution of drug resistance~\cite{hermsen2012rapidity,greulich2012mutational,Hermsen2016}. 
Greulich et al.~\cite{greulich2012mutational} considered a population adapting to a smooth gradient, which gradually lowers the growth rate of susceptible individuals. 
Hermsen et al.~\cite{hermsen2012rapidity} studied resistance evolution in a series of 
sharp step-like increases in concentration, where a novel resistance mutations was necessary for survival in the next step (the "staircase" model);  Hermsen~\cite{Hermsen2016} later proposed a generalization of the staircase model to continuous gradients. 
These previous studies focused on the speed of adaptation, i.e., how quickly the population evolves to tolerate high concentrations of antibiotics. 
In the context of the emergence of drug resistance, however, this observable alone ignores a crucial reality of antibiotic treatment: efficient drug treatment first and foremost aims to kill as many bacteria as possible, while limiting the rise of resistance mutation~\cite{Song2003,Ankomah10062014}. How this apparent trade-off can be optimized for populations in spatial gradients to prevent the evolution of drug resistance has so far been unexplored.  

Here, we present simulations, rationalized by a comprehensive analytical framework, of populations evolving resistance in a variety of spatial antibiotic concentration gradients and under the influence of convection. We measure the establishment probability of individual resistant mutants arising in a region occupied by susceptible wild type and find that successful, "surfing", mutations arise in a localized population patch close to the population front, the size of which depends on the relative strength of bacterial diffusion, antibiotic gradient steepness, and convection. We find that shallow gradients and flow towards higher antibiotic concentrations promote wild-type death, at the cost of increasing the establishment probability of resistance mutations. Conversely, populations in steep gradients and subject to flow towards lower antibiotic concentrations give rise to fewer drug-induced wild-type deaths but also produce fewer resistance mutants. We introduce the notion of a treatment efficiency, which quantifies this inherent trade-off between adaptation and death, and find it to be strongly modulated by gradient steepness and convection.

\section{Definition of the model}

\begin{figure}
\centering
\includegraphics[width=0.49\textwidth]{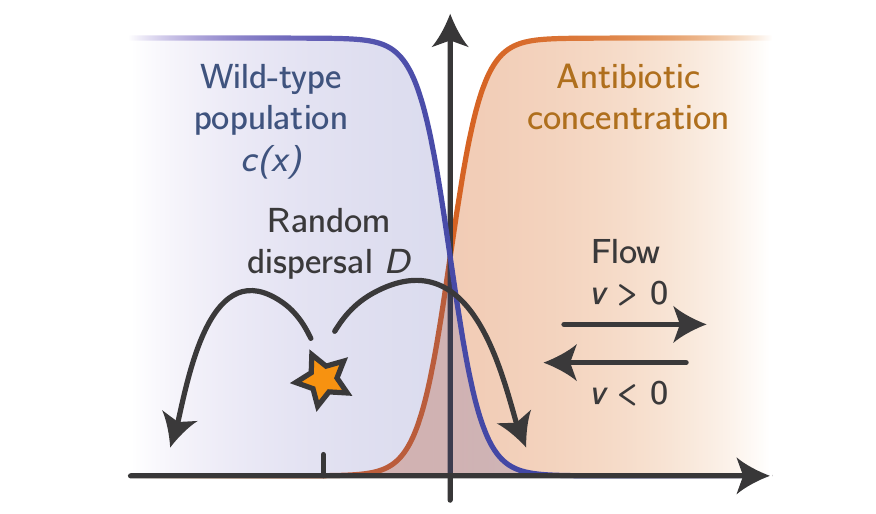}
\caption{Sketch of our modeling setup. We assume that, initially, a purely wild-type population has reached a steady-state density profile (blue) in the presence of a steady-state antibiotic concentration gradient (orange). Resistance mutations occur spontaneously in randomly drawn individuals and disperse, proliferate and die until extinction or ultimate fixation. Convection can either drive the population towards the antibiotic (co-flow, $v>0$) or away from it (counter-flow, $v<0$). Our goal is to analyze the establishment probability of resistance mutation for a given rate of wild-type killing.}
\label{fig:sketch}
\end{figure}

Consider a population in a fixed antibiotic concentration gradient. Where the antibiotic concentration is too high, the population cannot survive unless it evolves resistance to the antibiotic, and thus at steady-state, a population density profile $c(x)$ will develop. For an individual mutation conferring antibiotic resistance that occurs at position $x$ in the population 
we define the probability $u(x)$ that this mutation will be successful, i.e., establish first locally and eventually colonize areas where the antibiotic concentration is too high to allow for growth of the wild type. \hl{In doing so, we implicitly assume a very low mutation rate and neglect clonal interference to focus on the fate of individual mutations (SI section 6).} We come back to this question in the Discussion.

\hl{A drug gradient arises upon administering an antibiotic that is introduced 
at one location in the system (the source) and flows out of the system at another (the sink). The antibiotic may be influenced by convection, e.g., through blood flow or peristaltics in the gut, and be subject to degradation. At steady-state, a concentration gradient between source and sink is established, which may take a range of shapes, from a sharp, step-like gradient for strong convection, to a shallow gradient over the whole system size. Using a simple reaction-diffusion model, we show in the Supplementary Information (SI section 1) how the gradient depends on the distance between source and sink, the diffusivity of the antibiotic, and the speed of convection (see SI). To keep the discussion general, here we 
approximate the features of a typical antibiotic gradient by modeling it as a sigmoidal function that changes over a characteristic length scale $\lambda$, which sets
the gradient steepness, without trying to associate a particular value of $\lambda$ with a specific combination of real-life parameters.}

\hl{The wild-type population growing in this gradient will in general also be subject to convection with flow speed $v$. However, it is important to note this flow speed may be very different from the convection that affects the antibiotic gradient. As an example, the flow speed inside a blood vessel depends on the distance from the wall, and thus, surface-bound populations such as biofilms may experience much lower flow speeds than antibiotic in the bulk fluid. More generally, it is to be expected that fluid flow differentially affects bacterial cells and antibiotic molecules due to their different sizes, potential porosity of the surroundings, adhesion effects, etc, such that convective flows are arguably the rule rather than the exception. Therefore, without loss of generality, we may ignore the details of how a particular steady-state antibiotic gradient is generated, and focus on how the combination of antibiotic gradient and convection shapes the emergence of resistance.}

We model the effect of the antibiotic by a drug-induced death rate $b(x)$ of susceptible wild-type individuals giving rise to a net growth rate $s(x)$ of the wild type that ranges from the maximal growth rate $a_0$ (in antibiotic-free regions) to some negative net growth rate (where the antibiotic concentration is high). For simplicity, we model the death rate such that at the highest concentration wild-type individuals typically die within one generation. Given a net growth rate profile, we can compute the steady-state wild-type population density $c(x)$, whence we obtain the number $B$ of drug-induced wild-type deaths per generation, 
\begin{equation}
\mathrm{total\ death\ rate\ } B=\sum c(x) b(x).
\label{eq:B}
\end{equation}
The total death rate $B$ quantifies the efficacy of the antibiotic; it corresponds to the rate at which the population is hindered in its growth by the antibiotic. \hl{Note that since we only consider steady-state population densities and thus do not explicitly simulation the eradication of the wild-type population, our analysis relies on the immune system or slowly progressing antibiotic concentration gradients to completely eradicate the wild-type population~\cite{Ankomah10062014}. }

To quantify the emergence of resistance, we  measure the local mutant establishment probability $u(x)$. Since the probability that a mutation occurs in the first place is proportional to the wild-type population density $c(x)$, it follows that successful mutants can only arise where both the wild-type population density and the establishment probability are high (see Fig.~\ref{fig:cofxanduofx}). A measure for how readily new resistant mutants establish is thus given by the product of wild-type population density and the establishment probability~\cite{Hallatschek2008}, summed over the whole population~\cite{lehe2012rate}. We call this measure the \est, $R$,
\begin{equation}
\mathrm{\est}\ R= \sum c(x) u(x).
\label{eq:R}
\end{equation}
The \est, $R$,  is proportional to the rate at which new resistance mutations arise (at a low mutation rate $\mu$, see SI section 6) and establish in the population.  Alternatively, the \est, $R$, can be understood as a measure proportional to the mean establishment probability $\sum c(x)u(x)/\sum c(x)$, i.e., the probability that a mutation arising anywhere in the population establishes.

Finally, we define the treatment efficiency $Q$ as the ratio of the effective reduction in growth and the rate of adaptation to the antibiotic: a treatment is deemed particularly efficient if it can reduce the growth of the wild-type population while at the same time hindering the emergence of resistant phenotypes as much as possible. In our model, this corresponds to defining
\begin{equation}
\mathrm{treatment\ efficiency\ } Q=\frac{B}{R}.
\label{eq:Q}
\end{equation}
As we shall see below, the \est, $R$, and the total death rate $B$ typically follow the same trends, e.g., they are both larger in shallow gradients than in steep ones, as may perhaps be suspected intuitively. By contrast, the treatment efficiency $Q$ will turn out to be less accessible to intuition and require a detailed understanding of the population density and establishment probability profiles. In the following, we first present simulation results and then turn to analytical theory to rationalize our findings.

\section{Simulation results}

\begin{figure}
\centering
\includegraphics[width=\textwidth]{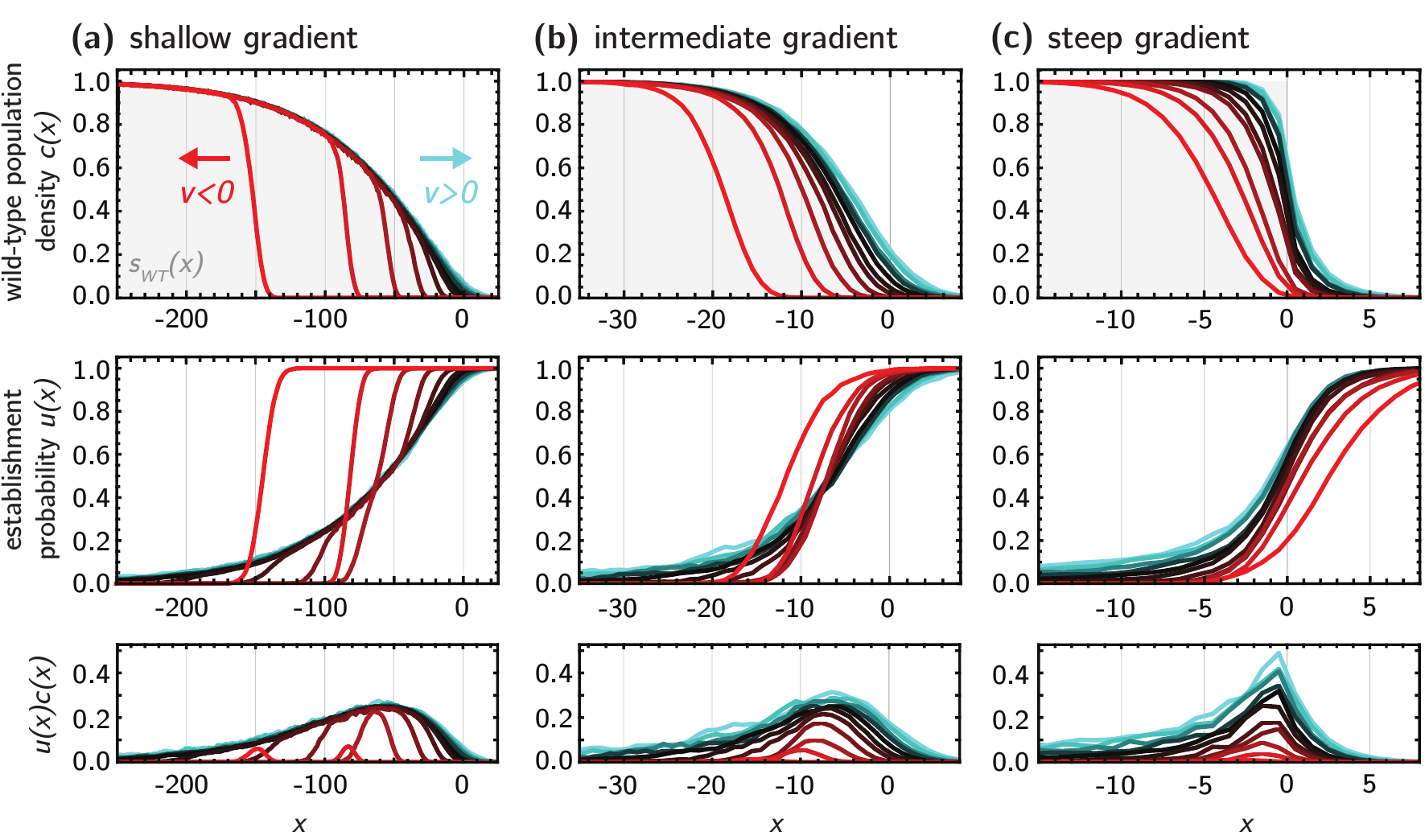}
\caption{Population density $c(x)$ (top row) and establishment probability $u(x)$ (middle row) for three different concentration profiles ($s_{WT}$, gray background; (a) shallow gradient, $\lambda=100$; (b) intermediate gradient $\lambda=10$; (c) step) and different values of $v=-0.4$ to $0.6$ ($v<0$, red tones; $v=0$, black; $v>0$, cyan tones), from stepping stone simulations. 
The bottom row shows the product $u(x)c(x)$, which identifies the localized region where successful mutants arise.}
\label{fig:cofxanduofx}
\end{figure}

\begin{figure}
\centering
\includegraphics[width=0.49\textwidth]{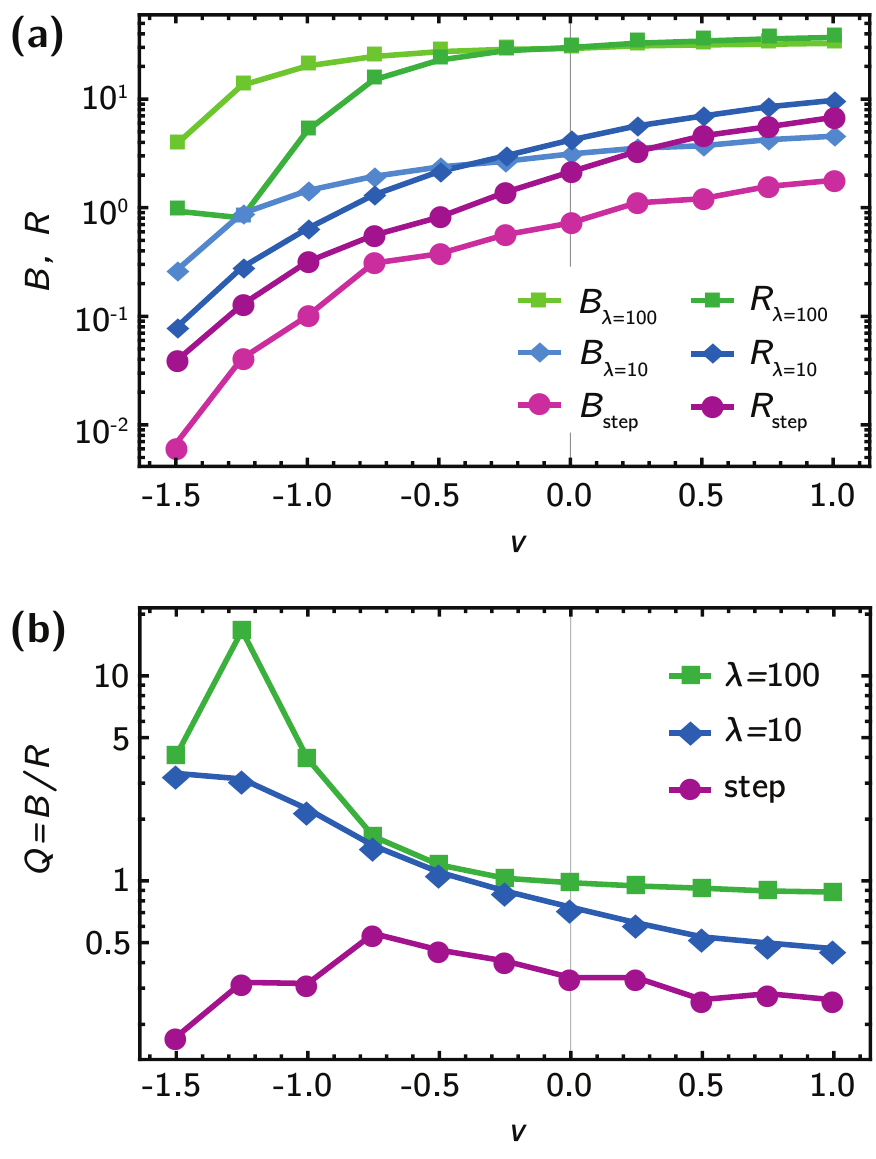}
\caption{Quantifying the trade-off between killing of wild-type cells and establishment of resistant mutants. (a) the total death rate $B$ and the \est, $R$,  for different antibiotic gradients (shallow gradient, $\lambda=100$, green; intermediate gradient, $\lambda=10$, blue; step, purple). Counter-flow ($v<0$) leads to drastic decrease of both $B$ and $R$, while co-flow has a milder effect, in accordance with the individual population and establishment probability profiles in Fig.~\ref{fig:cofxanduofx}. (b) The treatment efficiency $Q$, defined as the number $B$ of drug-induced deaths per generation divided by the \est, $R$,  (see eq.~\ref{eq:Q}). Counter-flow can increase the treatment efficiency by an order of magnitude in shallow gradients because $R$ is reduced by the effect of convection on both $u(x)$ and $c(x)$, while $B$ only captures changes in $c(x)$.}
\label{fig:RoverBandRandB}
\end{figure}

We simulate a population of wild-type individuals on a lattice of $L$ demes, where
each individual can migrate into a neighboring deme, replicate, and die, using a Gillespie algorithm~\cite{gillespie1976general,gillespie1977exact} (see \textit{Methods}). Wild-type population growth is limited to a carrying capacity $K$ by a logistic death term~\cite{Murray2002}. The population is subject to convective flow with speed $v$ and set in a one-dimensional antibiotic gradient (see the sketch in Fig.~\ref{fig:sketch}), which interpolates between maximal and zero antibiotic concentration over a characteristic length scale $\lambda$. The antibiotic gradient induces a wild-type death rate $b(x)$, giving rise to an effective growth rate $s(x)$ for the wild-type. For simplicity, we choose the functional form
\begin{equation}
s(x)=-a_0 \tanh\left(x/\lambda\right),
\label{eq:tanh}
\end{equation}
such that wild-type growth is strongly inhibited for $x\gg 0$ and proceeds unhindered with growth rate $a_0$ for $x\ll0$.

Following the equilibration to the steady-state profile $c(x)$ of the wild type, a single resistant mutant is inserted into the population at position $x$. The resistant mutant has the same birth rate as the wild type and is subject to the same carrying capacity as the wild-type, but it does not suffer from an increased death rate due to the antibiotics. We follow the mutant clone until it either goes extinct or reaches the far end of the simulation box, in which case we consider the mutant \textit{established}. The establishment probability $u(x)$ is then equal to the fraction of simulations in which a mutant introduced at $x$ managed to establish. 

Fig.~\ref{fig:cofxanduofx} shows the resulting population density $c(x)$ (top row) and establishment probability $u(x)$ profiles, for three different gradients (columns). Within each panel, different colors represent the profiles resulting under different flow speeds. \hl{Here, we adopt the usual convention of positive flow speed pointing to the right, which in our case corresponds to flow towards high antibiotic concentrations. We will refer to such flow as co-flow, and represent it graphically in hues of cyan. Conversely, negative flow speeds point to the right, towards lower antibiotic concentration; we term this counter-flow and use red tones.} 

In the absence of convection, shown in black, the wild-type population density $c(x)$ roughly follows the net growth rate $s_{WT}(x)$ (gray dotted line), and the establishment probability $u(x)$ is generally high where $c(x)$ is low, and vice-versa, since the mutants compete for resources with the wild-type. 

Convection affects the wild-type population density profile as intuitively expected, by stretching or compressing the population spatial range for co-flow and counter-flow, respectively. Interestingly, while counter-flow significantly alters the profile, the effects of co-flow are hindered by the drug profile, which prevents the wild-type from growing too far into the antibiotic region.

The effect of convection on the establishment probability is more complex, due to the generation of two competing processes that either help or hinder the mutant success. 
On the one hand, the changes in the wild-type profile described above alter the competition with the mutants, reducing the establishment probability in co-flow conditions, and increasing it in counter-flow conditions.  
On the other hand, mutants are also transported towards the direction of flow: their success rate is thus increased by co-flow and reduced by counter-flow. The presence of these two opposing forces generates interesting cut-offs along the profile, which we characterize in the Theory section.

The relative strength of the two competing effects can be quantified by the product $u(x)c(x)$, which represents the probability density of \textit{successful} mutants (bottom row in Fig.~\ref{fig:cofxanduofx}). Here we see that co-flow widens the probability peak, while counter-flow shrinks it and moves it away from the antibiotic region. This suggests that transportation of the mutants in the direction of flow rather than competition with the wild-type is the dominant effect in the mutant success rate. Since the area under these curves is the \est, $R$,  defined in eq.~\ref{eq:R}, we find that both the \est and the total death rate $B$ are higher in co-flow conditions and shallow gradients and lower in steep gradient with counter-flow, as shown in Fig.~\ref{fig:RoverBandRandB}a. Although $R$ and $B$  follow the same rough trends with flow speed $v$, the \est, $R$,  is generally more strongly affected by flow than the total death rate $B$. This is because flow alters both $c(x)$ and $u(x)$, but leaves the local death rates $b(x)$ unchanged.

This has dramatic effects on the treatment efficiency $Q=B/R$, shown in Fig.~\ref{fig:RoverBandRandB}b. Shallow gradients are generally conducive to a larger treatment efficiency $Q$ than steep gradients. In particular, strong counter-flow conditions ($v<0$) in shallow gradients give rise to very small $R$ compared to $B$, such that the treatment efficiency can be more than an order of magnitude larger than in the no-flow case. In steep gradients, although both $R$ and $B$ change over two orders of magnitude for different flow speeds, the treatment efficiency $Q$ remains roughly constant because both $R$ and $B$ are equally affected by flow (Fig.~\ref{fig:RoverBandRandB}a, purple lines).

In summary, our simulations show that the population density profile $c(x)$ as well as the local establishment probability $u(x)$ of resistance mutants are both strongly influenced by environmental parameters, in particular, by the steepness of the antibiotic gradient and the strength of convection. In the following, we lay out a mathematical model based on the theory of branching random walks that reproduces our key findings and elucidates the relative roles of gradient steepness and flow speed; detailed calculation are mostly relegated to the Supplementary Information.

\section{Theory}
\subsection{General framework}
\begin{figure}
\centering
\includegraphics[width=0.49\textwidth]{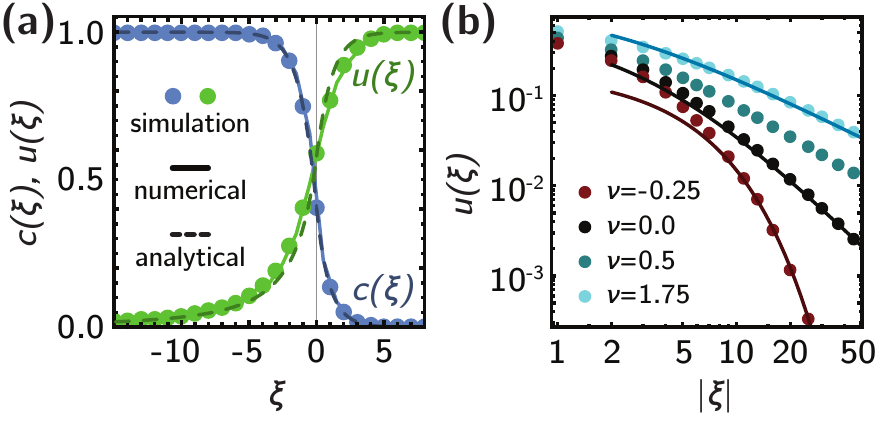}
\caption{The analytical model is in good agreement with the simulation results. (a) Population density $c(\xi)$ (blue) and establishment probability $u(\xi)$ profiles for a step-like antibiotic profile without external flow. Symbols are simulations, solid lines are numerical solutions to eqs.~\ref{eq:cofxandt} and~\ref{eq:survival-probability0}, dashed lines are analytical solution, eqs.~S7 and~S9. (b) Co-flow ($\nu > 0$, cyan) gives rise to broader algebraic establishment probability profiles, while counter-flow ($\nu<0$) gives rise to an exponential decrease. Both cases are asymptotically captured by the analytical approximations (solid lines, eqs.~S9, S15 and S16. } 
\label{fig:driftCU}
\end{figure}

We assume that the wild-type population density $c(x,t)$ (rescaled by the carrying capacity $K$) is described by the  reaction-diffusion equation~\cite{Nelson1997}
\begin{equation}
\label{eq:wt-profile}
\partial_t c(x,t) = D\partial_x^2 c -v \partial_x c +  s_{WT}(x)c-a_{WT}(x)c^2,\label{eq:cofxandt}                       
\end{equation}
where $a_{WT}(x)$ is the local wild type birth rate, $b_{Wt}$ is the local antibiotic-induced death rate, $s_{WT}(x)=a_{WT}(x)-b_{WT}(x)$ is the local net growth rate of the wild type, and $v$ is the external flow speed, representing, e.g., blood flow. This model is a straight-forward generalization for the standard Fisher model~\cite{fisher1937wave} to spatially inhomogeneous growth rates. Our model ensures that the steady-state local population density $c_{SS}$ depends explicitly on the local death rate $b_{WT}(x)$ when the death and birth rate profiles change sufficiently slowly in space, 
\begin{equation}
c_{SS}(x)=1-\frac{b_{WT}(x)}{a_{WT}(x)}.
\end{equation}
Our model, like the original Fisher equation, ignores the discrete nature of individuals, which has been shown to significantly alter the tip of the population front~\cite{Brunet1997,Hallatschek2011}, especially when the carrying capacity is small~\cite{Hallatschek2009} (see SI section~5 for a detailed discussion). Nevertheless, we expect good agreement between our model and simulations in terms of the total death rate $B$ and the \est, $R$, , whose values do not significantly depend on the population profile at the tip of the front. 

Given a single resistance mutation 
arises in the population at position $x$, 
its probability $u(x,t)$ 
to survive for a time $t$ can be derived by modeling the mutant lineage as a branching random walk; it obeys a nonlinear reaction-diffusion equation (see Ref.~\cite{lehe2012rate} and SI section 2),
\begin{equation}
  \label{eq:survival-probability0}
  \partial_t u(x,t)=D\partial_x^2u+v\partial_xu+s_{MT}(x) u-a_{MT}(x)u^2,
\end{equation}
where $a_{MT}(x)$ is the local birth rate of the mutants and $s_{MT}(x)=a_{MT}(x)-b_{MT}(x)$ is their net growth rate. The establishment probability $u(x)$, i.e., the ultimate survival probability for a mutation born at position $x$, as measured in simulations, is given by the steady-state solution of eq.~\ref{eq:survival-probability0}.

To mimic the situation in our simulations, we assume that the birth rate of wild-type and mutant is identical and constant, $a_{WT}(x)=a_{MT}(x)=a_0$, while the wild-type drug-induced death rate $b_{WT}(x)$ ranges from $-a_0$ to $a_0$. This implies that effect of the antibiotic is to increase the death rate of the wild type, while the drug-induced death rate of the resistant mutant is zero. The effective growth rate of the mutants is thus determined purely through competition with the wild type, i.e., $s_{\mathrm{MT}}(x) = a_0[1-c(x)]$. To model random dispersal and external flow, we have included diffusion and convection terms in eqs.~\ref{eq:cofxandt} and~\ref{eq:survival-probability0} (note the difference in sign between the convection terms). To get a feel for solution to the set of equations~\ref{eq:cofxandt} and~\ref{eq:survival-probability0}, we first study the case $v=0$ before turning on convection ($v\neq 0$).

\subsection{No flow}
We begin by considering the simplest functional form for the antibiotic gradient -- a step-like increase in concentration at $x=0$ that gives rise a net growth rate of $a_0$ for $x<0$ and $-a_0$ for $x>0$, i.e., 
\begin{equation}
s_{MT}(x)=a_0 \left[1-2\Theta(\xi)\right].
\end{equation} 
Such a sharp gradient could emerge, for instance, at the boundary of different tissues or organs with different affinities to store antibiotics~\cite{Nix1991,fu2015spatial}. Upon rescaling the spatial coordinate by the characteristic length scale $\ell=\sqrt{D/a_0}$, which can be intuitively understood as the typical distance that a mutant individual travels through random dispersal before replicating, eq.~\ref{eq:cofxandt} for the wild-type population density in this case becomes
\begin{equation}
0=\partial_\xi^2c+[1-2\Theta(\xi)] c-c^2
\label{eq:c-step}
\end{equation}
where $\xi=x/\ell$. Its solution, given analytically in SI section 3, transitions quickly from 1 to 0 over a distance $\sim 2\ell$ (Fig.~\ref{fig:driftCU}a). 

To find the establishment probability $u(\xi)$ far from the transition, we solve  
\begin{equation}
0=\partial_\xi^2u+[1-c(\xi)]u-u^2,
\label{eq:u-step}
\end{equation}
which, given the exact solution $c(\xi)$, can simply be integrated numerically (see \textit{Methods}). To make analytical progress, we approximate the wild-type population density with a step, i.e., $c(\xi)\approx \Theta(-\xi)$ and ask for solutions to eq.~\ref{eq:u-step} far from the transition region. For $\xi>0$, $u(\xi)$ also approaches 1 exponentially quickly over a distance $\ell$, since the lack of competition with the wild-type in this region facilitates the establishment of resistance mutations. Instead, far inside the bulk of the population ($\xi<0$), we find that the establishment probability exhibits a long tail, $u(\xi)\sim \xi^{-2}$ (Fig.~\ref{fig:driftCU}a). The long tail implies that mutations arising deep inside the bulk of the population still have a relatively high chance of establishing, which explain the asymmetry of the simulation curves in Fig.~\ref{fig:cofxanduofx}c.

For gradients varying over length scales longer than $\ell$, we model gradients decaying over a characteristic length scale $\lambda$ with a net growth rate $s(\xi)$ of the form  eq.~\ref{eq:tanh}. Introducing the steepness parameter $m=\ell/\lambda$, which quantifies the relative length scales associated with diffusion and drug gradient, we can write the steady-state equation for the wild-type population density as
\begin{equation}
0=\partial_\xi^2c+\tanh(-m \xi) c-c^2.
\label{eq:c-shallow}
\end{equation}
While eq.~\ref{eq:c-shallow} does not have an analytical solution, we can change variables to $\xi'=\xi m$ to see that the diffusion term in eq.~\ref{eq:c-shallow} can be neglected when $m\ll 1$ and thus $c(\xi)\approx s_{WT}(\xi)$. More generally, for mutants in shallow gradients with a mutant net growth rate $s_{MT}(\xi)$ (see eq.~\ref{eq:survival-probability0}), we get the eventual establishment probability of the familiar form
\begin{equation}
u(\xi)=\frac{s_{MT}(\xi)}{a_{MT}(\xi)}.
\label{eq:quasi-static}
\end{equation}
This so-called quasi-static approximation is a straight-forward extrapolation of the well-mixed result and has been used to model the establishment probability by Greulich et al.~\cite{greulich2012mutational}.

Once solutions for $c(\xi)$ and $u(\xi)$ are found, they can be used to compute the \est, $R$,  and total death rate $B$. 
We find asymptotically
\begin{equation}
R \approx \left\{\begin{array}{lr}
         1.91\ell, & \textrm{for }\lambda \ll \ell,\\
       (1-\ln 2)\lambda , & \textrm{for } \lambda \gg \ell,
        \end{array}\right.
        \label{eq:B-analytical}
\end{equation}
and 
\begin{equation}
B \approx  \left\{\begin{array}{lr}
         0.38\ell, & \textrm{for } \lambda  \ll \ell 
         ,\\
      (1-\ln 2)\lambda, & \textrm{for } \lambda \gg \ell
       ,
        \end{array}\right.
        \label{eq:R-analytical}
\end{equation}
which agrees well with the numerical result 
in the two limiting cases $\lambda \ll \ell$ and $\lambda \gg \ell$ (Fig.~\ref{fig:BRm}). Thus, the treatment efficiency $Q$ is constrained to a relatively small range in the absence of convection.

\subsection*{Co-flow and counter-flow}
\begin{figure}
\centering
\includegraphics[width=0.49\textwidth]{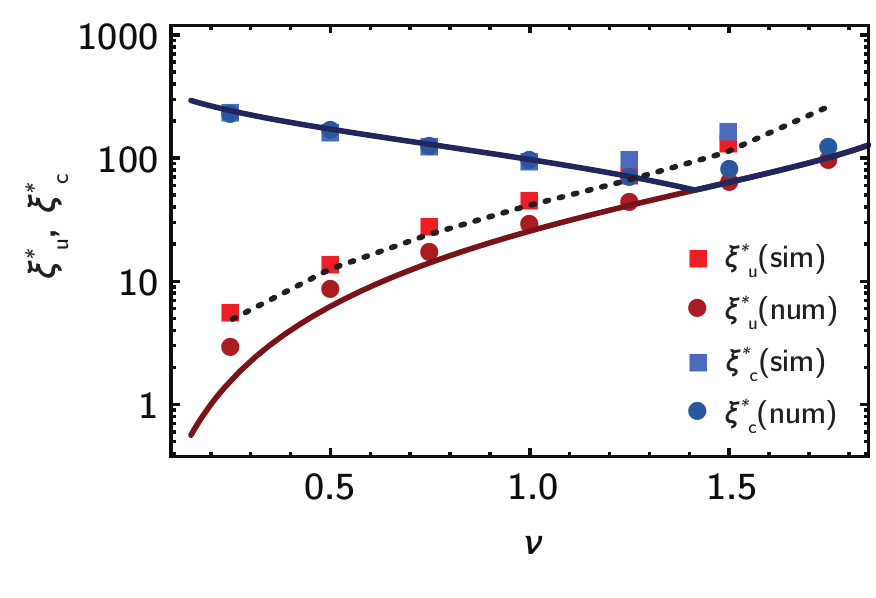}
\caption{In shallow gradients, counter-flow gives rise to sharp cut-offs in the population density ($\xi^*_c$) and establishment probability ($\xi^*_u$) profiles (see Fig.~\ref{fig:cofxanduofx}a). The cut-off  positions are well-captured by the analytical approximations (solid lines, eqs.~\ref{eq:cut-offc} and~\ref{eq:cut-offu}). Squares are cut-offs extracted from simulation, dots and dotted line from numerical solutions to eq.~\ref{eq:cofxandt} and~\ref{eq:survival-probability0} without and with a growth rate cut-off (SI section~5).
} 
\label{fig:driftcutoffs}
\end{figure}

We now investigate the influence of convection on the population density $c(x)$, the establishment probability $u(x)$, and finally, the treatment efficiency $Q$, which, in the absence of convection, is naturally constrained to a relatively small range, see eqs.~\ref{eq:B-analytical} and~\ref{eq:R-analytical}. We observed in simulations (Fig.~\ref{fig:cofxanduofx}) that convection will incur either a depletion or enrichment of both mutant and wild types in the antibiotic region, depending on whether the convection points towards higher or lower antibiotic concentrations (co- or counter-flow, respectively). To study the effects of convection analytically, we set $v\neq 0$ in eqs.~\ref{eq:cofxandt} and~\ref{eq:survival-probability0}, and introduce the dimensionless flow speed $\nu=v/\sqrt{Da_0}$. The characteristic flow speed $v_c$ is closely related to the Fisher wave speed $v_F$, which corresponds to the expansion speed of a freely growing population. If convection is too strong, i.e., the flow speed is great than the Fisher wave speed, ($|\nu|>2$), no steady-state population density exists because the population is washed away (see SI section 5), and we are therefore constrained to $|\nu|<2$.

In step-like gradients ($m \gg 1$), even in the presence of convection, the population density $c(\xi)$ approaches its asymptotic values exponentially fast, as can be seen by expanding eq.~\ref{eq:cofxandt} to first order around its fixed points and solving the resulting linear differential equations (see SI section 4). As in the no-flow case, we can therefore approximate $c(\xi)$ with a step function to find approximate solutions for $u(\xi)$ far from the step gradient. With co-flow ($\nu>0$), the diffusion term in eq.~\ref{eq:survival-probability0} can be neglected to first order and hence the establishment probability has a very broad tail, $u(\xi)\sim (1+\xi/\nu)^{-1}$ (Fig.~\ref{fig:driftCU}b, SI eq. S15). With counter-flow ($\nu<0$), the diffusive term is essential because the convection term cannot balance the non-linearity since both are negative. This gives rise to a rapid exponential decay $u(\xi)\sim e^{-\nu \xi}$ (Fig.~\ref{fig:driftCU}b).

In shallow gradients ($m \ll 1$), co-flow has little effect on the population density (and correspondingly establishment probability) profiles because the convection term $\nu c'(\xi)$ in eq.~\ref{eq:cofxandt} is roughly proportional to $m \ll 1$. In counter-flow condition, our model reproduces the cut-off in the population density $c(\xi)$ seen in our simulations, which traces the net growth rate $s_{WT}(\xi)$ until it drops sharply. The position of this cut-off can be estimated from the definition of the Fisher wave speed $\nu_F$, which depends on the wild-type net growth rate $s_{WT}(\xi)$ (see also SI section 5). In shallow gradients, $s_{WT}(\xi)$ changes little around the cut-off, such that the (rescaled) Fisher wave speed becomes position-dependent $\nu_F(\xi)=2\sqrt{s_{WT}(\xi)/a_0}$. Hence, no wild-type population density can be maintained in regions where the flow speed $\nu$ becomes larger than the local Fisher wave speed $\nu_F(\xi)$, and solving for $\xi$ gives the position of the cut-off 
\begin{equation}
\xi^*_{c}= -\frac{1}{m}\rm{arctanh}(\nu^2/4).
\label{eq:cut-offc}
\end{equation}
Similarly, we find two cut-offs in the establishment probability profile $u(\xi)$: since in shallow gradients, $u(\xi)\approx 1-c(\xi)$, $u(\xi)$ approaches 1 sharply when $c(\xi)$ drops to zero at $\xi^*_c$. In addition, since the mutants have a local net growth rate that depends on the wild-type density, their speed limit is by $\nu_F(\xi)=2\sqrt{1-c(\xi)}$, which gives another cut-off at the position 
\begin{equation}
\xi^*_{u}= -\frac{1}{m}\rm{arctanh}(1-\nu^2/4).
\label{eq:cut-offu}
\end{equation}
When $\nu <-\sqrt{2}$, the two cut-offs reduce to one, 
and both $c(\xi)$ and $u(\xi)$ exhibit a sharp transition at $\xi^*_{c}$ (see Figs.~\ref{fig:cofxanduofx}a). 

\begin{figure}
\centering
\includegraphics[width=0.49\textwidth]{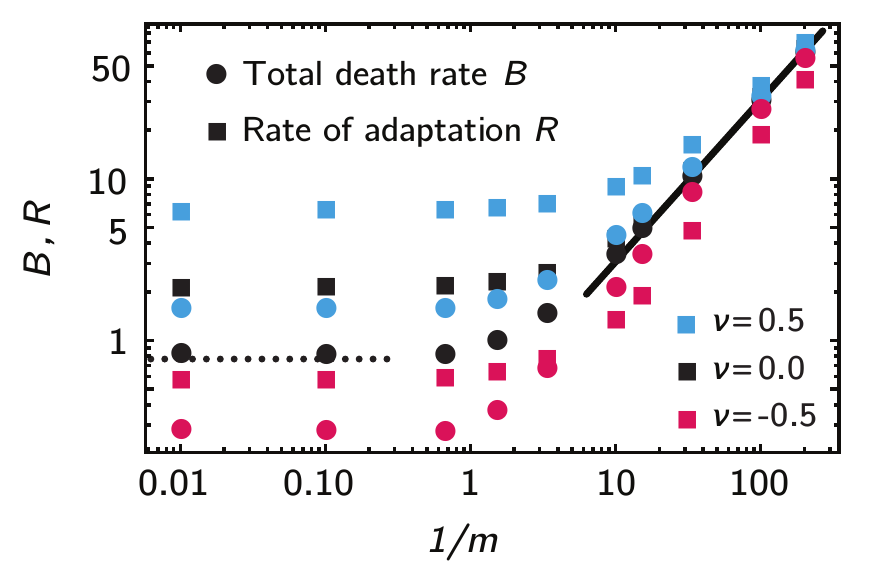}
\caption{Total death rate $B$ and \est, $R$ (symbols, from numerical solutions to eq.~\ref{eq:cofxandt} and~\ref{eq:survival-probability0}),  vary as a function of the rescaled gradient length $\lambda/\ell=1/m$. In steep gradients ($m\gg 1$), $B$ and $R$ approach constants. In the limit of shallow gradients ($m \ll 1$), both $R$ and $B$ are proportional to $1/m$. In the absence of external flow, both limits are captured by the analytical results, eqs.~\ref{eq:B-analytical} and~\ref{eq:R-analytical}.}
\label{fig:BRm}
\end{figure}

Fig.~\ref{fig:driftcutoffs} compares the cut-off positions in the establishment probability profiles from the numerical model and simulations. While there is excellent agreement between this theoretical predictions, eqs.~\ref{eq:cut-offc} and~\ref{eq:cut-offu}, and the numerical evaluation of the model, as well as with the cut-off $\xi^*_{u}$ of the simulated establishment probability $u(\xi)$, the cut-off $\xi^*_{c}$ of the simulated population density $c(\xi)$ appears shifted towards higher $\nu$. This effect can be traced back to number fluctuation at the front of the population where the population size is small, which leads to a significant shift in the cut-off depending on the carrying capacity $K$ (SI section 5, Fig.~\ref{fig:driftcutoffs}, black dotted line). 

\begin{figure}
\centering
\includegraphics[width=0.49\textwidth]{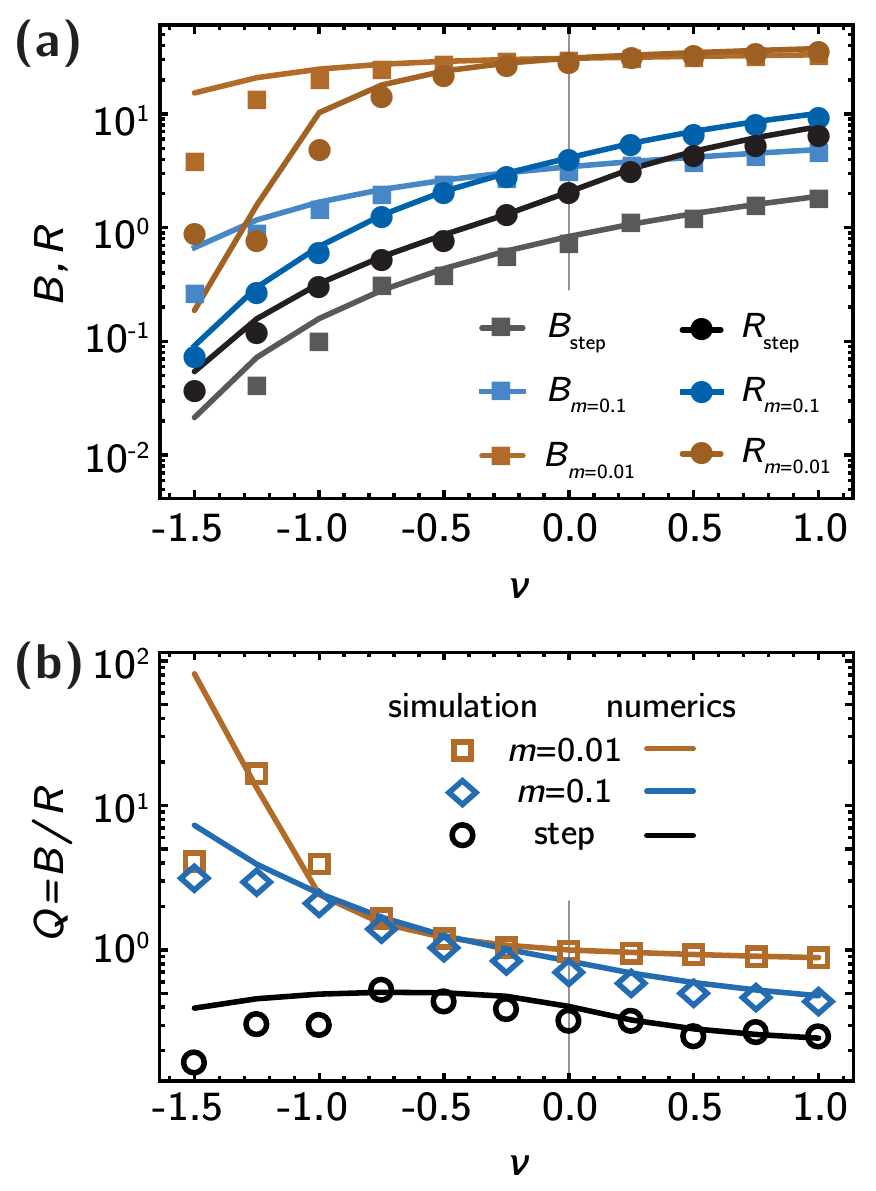}
\caption{Comparison between simulations and numerical solutions. (a) Total death rate $B$ and \est, $R$,  computed from simulations (symbols) and numerical solutions to eqs.~\ref{eq:cofxandt} and~\ref{eq:survival-probability0} (lines), in three different gradients (as in Fig.~\ref{fig:RoverBandRandB}). The agreement is very good, except for strong counter-flow, where number fluctuations at the front become important, as explained in the main text. (b) The treatment efficiency $Q$ found in simulations is also well-captured by our model, except for strong counter-flow.}
\label{fig:Qcomparison}
\end{figure}

By integrating the numerical solutions to eqs.~\ref{eq:cofxandt} and~\ref{eq:survival-probability0}, we can compute the total death rate $B$ and the \est, $R$,  and compare with the simulation results. As shown in Fig.~\ref{fig:Qcomparison}, our model reproduces the phenomenology of the simulation very well, except in strong counter-flow conditions, where the number fluctuations at the front give rise to mostly quantitative differences.

\section{Discussion}

\begin{figure}
\centering\includegraphics[width=0.49\textwidth]{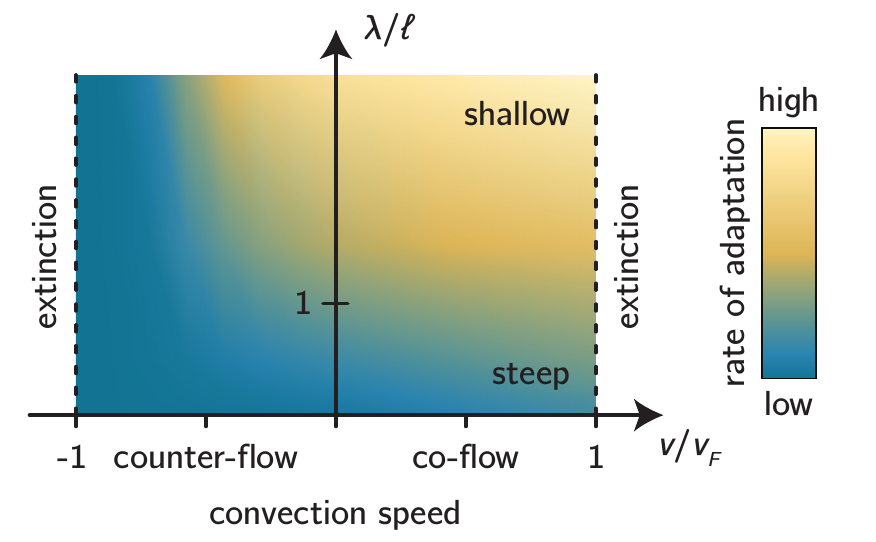}
\caption{\hl{Schematic summary of our main results. Identifying the characteristic length scale $\ell$ allows us to classify gradients as steep (bottom) or shallow (top), depending on how $\ell$ compares with the gradient length scale $\lambda$. Convection leads to extinction when it is too strong compared with the characteristic convection speed $v_F$, no matter whether convection is directed towards higher (co-flow) or lower (counter-flow) antibiotic concentration. Shown in the background is the rate of adaptation of the population to the antibiotic (blue colors, low adaptation; orange colors, high adaptation).}}
\label{fig:discussionfigure}
\end{figure}

We presented stochastic simulations and an analytical model to study the fates of individual resistance mutations in spatial antibiotic gradients, for the first time also including the effects of convection. Our analytical model allows us to identify two characteristic length scales, $\lambda$ and $\ell$, which, together with the growth rate $a_0$, determine the fate of resistance mutations (see Fig.~\ref{fig:discussionfigure}). The gradient length $\lambda$ describes the characteristic length scale over which the antibiotic concentration varies. Individuals travel a characteristic distance $\ell=\sqrt{D/a_0}$ through random dispersal with migration rate $D$ before replicating. The ratio $m=\ell/\lambda$ determines whether a gradient is shallow or steep. If the gradient is shallow ($\lambda \gg \ell$), the antibiotic concentration does not appreciably change over distances accessible to a single individual in one generation, such that the population is locally adapted to the antibiotic concentration and changes only on length scales of order $\lambda$.  If the gradient is steep ($\lambda \ll \ell$), an individual close to the gradient can travel between regions of high and low antibiotic concentration in its lifetime such that the population density is constant far from the gradient and changes relatively abruptly over a distance $\ell$ around the gradient. 

From the migration length $\ell$, we also identify the characteristic speed $v_c=\sqrt{Da_0}=\ell a_0$, which is closely related to the expansion speed $v_F=2v_c$ of a freely expanding population~\cite{fisher1937wave}. Hence, if convection is too strong, i.e., if the convection speed $|v| > v_F$, the population will not be able to grow against the flow and be simply washed away (Fig.~\ref{fig:discussionfigure}; this akin to the extinction transition of populations due to convection~\cite{dahmen2000life,geyrhofer2013stochastic}. For the emergence of resistance, we are thus naturally restricted to the region $|v| < v_F$.

It is illuminating to estimate the typical range of $\ell$ and $v_c$ for microbial communities. A typical (non-motile) bacterial cell may have a diameter of 1$\mu m$, swimming in a medium of viscosity comparable to that of water (e.g., blood~\cite{Rand1963}), which gives a diffusivity of order 0.1-1$\mu m^2/s$. The motion of motile bacteria is characterized by much larger diffusivities, up to tens or even hundreds of $\mu m^2/s$~\cite{kim1996diffusivity}. Together with a typical growth rate of $0.5-2\mbox{hr}^{-1}$, this gives a possible range for $\ell$ between 50$\mu$m and several millimeters. For comparison, in a microfluidic experiment by Zhang et al.~\cite{Zhang1764} the length scale on which the drug gradient varied was $\approx 200\mu$m such that $\lambda/\ell\sim1$. For another recent experimental study by Baym et al.~\cite{Baym2016} with a reported spreading velocity of 40mm/hr, we find $\ell\approx 1-20$mm. Thus, depending on the properties of the bacteria and the antibiotic gradient, both shallow ($\lambda \gg \ell$) and steep ($\lambda \ll \ell$) gradients can plausibly arise. \hl{Similarly, with our range of parameters for $D$ and $a_0$, we find typical values for $v_c \approx 0.1-10\mu m/$s, which is on the same order of magnitude as estimated  flow speeds in the gut, and achieved in an artificial gut microfluidic system~\cite{Cremer2016}}.

We have quantified the emergence of resistance by computing the rate of adaptation for a wide range of gradient steepnesses and convection speeds. As a general rule, our model predicts that the \est in shallow gradients is roughly proportional to the gradient length $\lambda$, as long as $\lambda$ is smaller than the system size. This is in agreement with previous results from Greulich et al.~\cite{greulich2012mutational}, who employed the quasi-static approximation, eq.~\ref{eq:quasi-static}. As we have shown, this approximation breaks down when $\lambda \approx \ell$. In the opposite limit, $\lambda \ll \ell$, i.e., in steep gradients, we find a finite rate of adaptation, in agreement with the so-called staircase model of Hermsen et al.~\cite{hermsen2012rapidity}. This is because mutants born in low antibiotic concentration regions can migrate into regions of high concentration, where they enjoy a big selective advantage. \hl{Our model serves as a generalization of these two previous model by capturing the full crossover from the "mutate and migrate" mechanism governing adaptation in steep gradients to the "local" adaptation in shallow gradients~\cite{Hallatschek2012}.}

\hl{Convection can have a momentous influence on the rate of adaptation in our model. Co-flow (convection towards higher antibiotic concentration) increases the rate of adaptation by extending the range of the wild-type population further into high concentration regions, where resistance mutants have a strong advantage. At the same time, co-flow transports resistant mutants born in low-concentration regions (where the wild-type population density is high) to the front. In the same manner, counter-flow (convection towards lower antibiotic concentration) can restrict the range of the wild-type population and prohibit resistance mutation from establishing when they arise far away from the antibiotic gradient. This can lead to a decrease of the rate of adaptation by several orders of magnitude for strong counter-flow, until eventually the population goes extinct when counter-flow becomes too strong (see Fig.~\ref{fig:discussionfigure}). }

In the context of antibiotic treatment of an infection, the eradication of the infecting bacterial population is paramount~\cite{Song2003} and therefore, we have argued that focusing on the rate of adaptation \textit{alone} may be misleading. To quantify the trade-off between wild-type killing and the emergence of resistance, we measured the reduction in pathogenic growth relative to the rate of adaptation. This measure, which we call the treatment efficiency, has an intuitive meaning: a high treatment efficiency implies a strong reduction in the growth of the bacteria before a resistance mutation arises and establishes. In our model, the treatment efficiency is strongly affected by counter-flow, where it can be an order of magnitude higher than in a no-flow scenario. \hl{This is ultimately a consequence of the rate of adaptation being doubly affected by convection, because convection alters both the wild-type population density (thus changing the local competition that resistant mutant clones face) and the dynamics of individual mutants, which are carried away from the high antibiotic concentrations they require in order to establish. Despite its intuitive meaning, however, the treatment efficiency as defined here makes no predictions about optimal treatment regimens; it merely serves as a quantification of the inherent trade-off between wild-type eradication and selection for resistance. }

Our model is restricted to scenarios where interference between multiple mutant clones can be neglected, since we follow the establishment of individual mutations.  Therefore, our model, operates exclusively in the "mutation-limited" regime~\cite{Hermsen2016}, where the rate of adaptation is dominated by the waiting time until the establishment of mutations. By contrast, both Hermsen~\cite{hermsen2012rapidity,Hermsen2016} and Greulich~\cite{greulich2012mutational} consider the establishment of many, potentially contemporaneous mutations. In very shallow gradients, this can lead to a "dispersion-limited" regime if the mutational supply is large; the rate of adaptation is then dominated by the speed with which established mutations invade previously uninhabitable territory. \hl{We derive upper limits for the mutation rate $\mu$ per generation for our model to be applicable (SI section 6) and find that in steep gradients, clonal interference is negligible as long as $\mu K \ll 1$, where $K$ is the typical population size in a population patch of size $\ell$. This result is similar to what is expected in well-mixed populations~\cite{Desai2007a} (see SI section 6). In shallow gradient, the corresponding condition is $\mu K \ll (\ell/\lambda)^2$. Thus, if the gradient is too shallow for a given mutation rate, mutations enter the population too fast and clonal interference effects are expected.} For resistance mutations with small target sizes, mutation rates can be very small, typically less than $10^{-6}$~\cite{Martinez2000,Woodford2007}, and thus our approximation may be accurate even for relatively large carrying capacities $K$. Even when the mutation rate is high, we expect convection and spatial gradients to have the same qualitative effects on the establishment of resistance mutations, and thus our results should remain qualitatively correct even beyond the single-mutation regime considered here. 

Our analytical model extends previous models~\cite{greulich2012mutational,hermsen2012rapidity,Hermsen2016} by incorporating arbitrary gradients and convection, and it can serve as a general branching process framework to study adaptation in spatially heterogeneous environments. For instance, we can easily incorporate antibiotic sanctuaries which are predicted to facilitate the emergence of resistance~\cite{kepler1998drug,fu2015spatial}. Other, more complex gradients may be necessary to describe cases in which the death rate depends on the antibiotic concentration on a non-linear manner~\cite{hermsen2012rapidity}. Our model can also easily be re-interpreted to apply to a broad range of different ecological scenarios, like heterogeneous nutrient concentration. To make our model more realistic, it would be useful to model the bacteria as having a finite size, such that the population front can advance through mere growth, even against strong counter-flow~\cite{Tesser2016}. For strong co-flow, individuals may also de-adhere and be carried away from the bulk population, thus founding extant colonies that enjoy large growth rates in the absence of competition for resources. Such processes can be studied by generalizing the diffusion term in our model to a long-range dispersal term as used frequently to model epidemics~\cite{Hallatschek2014}. Since long-range dispersal can allow mutants far from the population front to escape the bulk population, we expect it to increase the total establishment probability and thus the rate of adaptation relative to short-range dispersal as discussed here. Another interesting generalization of the model would be to extend the model to two-dimensional populations since real biofilms typically grow as two-dimensional communities, with complex spatial patterns. The establishment of beneficial mutations in microbial colonies has recently been discussed~\cite{Lavrentovich2013,Gralka2016}. Due to the particular strength of genetic drift at the front of such populations, beneficial mutations first have to reach a threshold size (depending on the strength of the selective advantage) neutrally before they become established. Once the mutant clone reaches the threshold size, the selective advantage of the mutants can deterministically drive them to fixation in the population. During the initial phase, the mutant clone is contained between boundaries with characteristic stochastic properties that are not captured in our one-dimensional model~\cite{Fusco2016}. However, if the threshold size is small, the boundary fluctuations will not have a large impact on the growth of mutant clones. In such cases, we expect our results to apply also to two-dimensional populations.

The emergence of drug resistance remains a topic of significant interest, both from a scientific and a public health point of view. Considerable effort is brought forward to create novel antibiotics~\cite{seiple2016platform} and new therapy strategies are developed that attempt to limit the emergence of resistance~\cite{Enriquez-Navas2016,Paterson2016}, but more research is needed to understand how resistance evolves in  complex spatio-temporal settings like the spatial gradients discussed in this paper. In particular, as we have shown here, convection constitutes an important factor in shaping the adaptation to antibiotics in spatial concentration gradients and should receive more attention from both theorists and experimentalists.

\section*{Acknowledgments}
Research reported in this publication was supported by the National Institute of General Medical Sciences of the National Institutes of Health under Award Number R01GM115851, by a National Science Foundation Career Award and by a Simons Investigator award from the Simons Foundation (O.H.). The content is solely the responsibility of the authors and does not necessarily represent the official views of the National Institutes of Health. The simulations were run on the Savio computational cluster resource provided by the Berkeley Research Computing program.

\section{Methods}
\subsection*{Individual-based simulation}
We perform individual-based, stepping stone simulations where both wild-type and mutants are modeled explicitly. The population is divided into demes with carrying capacity $K=100$ on a one-dimensional lattice. Wild-type and mutants replicate at a rate $a_0$ and migrate at rate $D$ independently on their position. Wild type in deme $x$ die at a rate $b(x)=1+\tanh(m x)$. Since we assume throughout that the antibiotic leaves the mutants unaffected,  mutants do not die in our simulations. Analogously to a Gillespie algorithm, in each simulation step, a birth, death or migration event is performed according to its relative rate~\cite{gillespie1976general,gillespie1977exact}, as follows. 

\begin{itemize}
\item \textbf{Birth.} Birth events occur at a total rate equal to $a_0\sum_x c(x)$, where $c(x)$ is the total number of individuals in deme $x$. For each birth event, a source individual is selected at random and replicated into a random target site between 1 and K within the same deme. Because the target site can either already be filled with an individual or be empty, this move effectively translates into logistic growth within the deme. 
\item \textbf{Death.} In our model, only the wild type can die, thus deaths have a total rate corresponding to $\sum_i b(x) c_\mathrm{wt}(x)$, where $c_\mathrm{wt}(x)$ represents the number of wild type individuals in deme $i$. To perform a death event, first, a deme $x$ is picked proportionally to its relative death rate  $b(x)c_\mathrm{wt}(x)$, and then, a random wild type within the same deme selected to be removed. 
\item \textbf{Migration.} Migrations are performed at a rate $D\sum_x c(x)$ by picking a random individual and swapping it with a randomly selected target site from one of the two neighboring demes. As in the case of birth events, the target site can either correspond to an individual, or to an empty site. 
\item \textbf{Time step.} Time is tracked by sampling a time interval $\delta t$ from an exponential distribution with rate $\sum_x\left[(a_0+D)c(x)+b(x)c_\mathrm{wt}(x)\right]$, as in a standard Gillespie algorithm. The total elapsed time is the sum of the sampled time intervals.
\item \textbf{Convection.} Convection (with convection speed $v$) is implemented by shifting the population by one deme away or towards the antibiotic gradient, for negative or positive convection respectively. The convection strength is controlled by performing the shift at a rate $1/|v|$, i.e., any time the time elapsed since the last shift is greater than $1/|v|$.   
\end{itemize}

For each simulation, we first allow the wild type to reach the steady-state profile $c(x)$. We then introduce one mutant element at position $x$ and run the simulation until either all mutants go extinct, or mutants reach the last deme in the simulation box. No further mutations are allowed in the course of the simulations. The probability of fixation $u(x)$ is then computed as the proportion of the simulations in which a mutant introduced at $x$ reached fixation.

\subsection*{Numerics}
Numerical solutions to eqs.~\ref{eq:cofxandt} and~\ref{eq:survival-probability0} were obtained by evaluating the differential equations using Mathematica's built-in \textit{NDSolve} routine with the backwards differentiation (BDF) method, with a maximum step size of 1 and a domain size of 1000. Initial conditions were chosen according to the analytical approximations to the steady-state profiles given in the text. This is done to speed up computation and increase numerical stability, but is otherwise inconsequential; starting with different initial conditions leads the same final solution. To obtain steady-state profiles, we solve the full time-dependent problems until the solution no longer changes for longer evaluation times.

To compute the establishment probability $u(x)$ in realistic wild-type population profiles $c(x)$, we first computed the steady-state population density and then used the final profile to compute the (constant in time) local death rate for the mutants $b(x)=1-c(x)$. The resulting numerical solutions were integrated numerically using Mathematica's built-in \textit{NIntegrate} routine to obtain the total death rate $B$ and the \est, $R$, .

\setcounter{section}{0}
\setcounter{figure}{0}
\renewcommand{\thesection}{S}
\renewcommand{\thesubsection}{S\arabic{subsection}}

\setcounter{subsection}{0}
\setcounter{equation}{0}

\section*{Supplementary Information}
\renewcommand{\thefigure}{S\arabic{figure}}
\renewcommand{\theequation}{S\arabic{equation}}
\renewcommand{\thetable}{S\arabic{table}}
\renewcommand{\thesection}{S\arabic{section}}
\renewcommand{\thesubsection}{S\arabic{subsection}}

\subsection{Antibiotic concentration profile}
\begin{figure}
\centering
\includegraphics[width=0.49\textwidth]{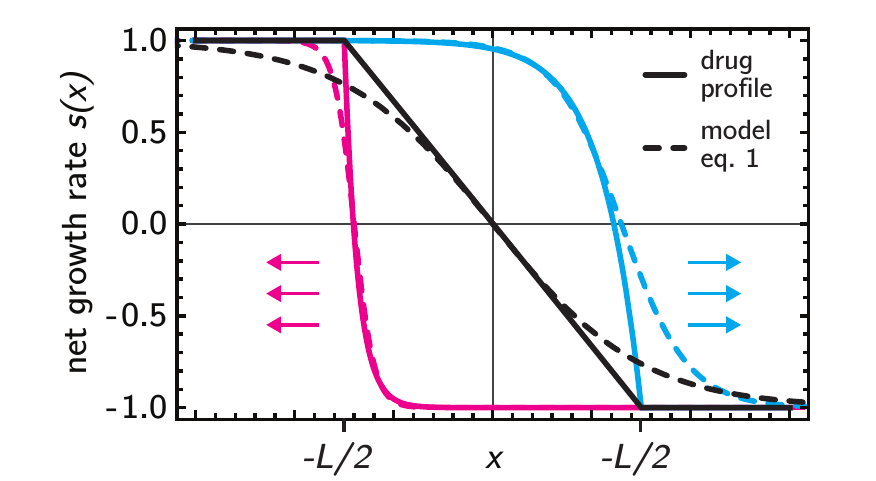}
\caption{Example of realistic growth rate profiles for different convection speeds $v_C$ ($v_C>0$, cyan; $v_C=0$, black; $v_C<0$, magenta). Solid lines are solutions to eq.~\ref{eq:SIconc}, dashed lines are fits with eq. 1 in the main text.}
\label{fig:abconc}
\end{figure}
In the main text, we suppose an antibiotic gradient varying over a characteristic length scale $\lambda$. Here, we present an example for how such an antibiotic gradient may arise. We use a simple reaction-diffusion model for the antibiotic concentration $C$ in a one-dimensional system of length $L$ to show that our model for the antibiotic gradient, eq.~1 in the main text, can approximately capture a variety of realistic gradient profiles. 

Consider an antibiotic source at position $-L/2$ and an antibiotic sink at position $L/2$. The antibiotic concentration at the source is $c_\infty$ and the antibiotic concentration at the sink is 0. The drug diffuses with diffusion constant $D_C$ and is subject to convection with speed $v_C$. Additionally, the antibiotic may be degraded, e.g., by cells metabolizing the drug or through chemical degradation, at a rate $\gamma$. Under these assumption, the steady-state antibiotic concentration profile is described by the  reaction diffusion equation
\begin{equation}
0=D_C\partial_x^2 C(x) - v_C \partial_x C(x) -\gamma C(x).
\label{eq:SIconc}
\end{equation}
The resulting net growth rate $s(x)= 1-2C(x)/C_{\infty}$ is shown for three values of $v_C$ and $\gamma=0$ in Fig.~\ref{fig:abconc} (solid lines). The dashed lines are approximations of the form given in eq.~1 in the main text. For our purposes, eq.~1 gives a good enough fit to (simple) realistic antibiotic profiles.

\subsection{Derivation of survival probability}
Let $u_x(t)$ denote the probability that a mutation born at lattice site $x$ survives  for time $t$. Denote by $a(x)$ and $b(x)$ the local birth and death rate, respectively, and let $v_+$ and $v_-$ be the rates to migrate a distance $\delta x$ (i.e., one lattice site) to the right and to the left, respectively. Then, the $u(x,t)$ after a short time interval $\epsilon$ satisfies the equation
\begin{eqnarray}
u_x(t+\epsilon)&=\epsilon a(x)\left\lbrace 1-\left[1-u_x(t)\right]^2\right\rbrace\nonumber \\
&+\epsilon\left\lbrace v_+ u_{x+\delta x}(t) + v_- u_{x-\delta x}(t)\right\rbrace \\
&+\left\lbrace 1-\epsilon\left[a(x)+b(x)+v_++v_-\right]\right\rbrace u_x(t).\nonumber
\end{eqnarray}
The first term on the right-hand side accounts for the fact that when the initial mutant divides then there are two mutants, and the probability of survival of at least one lineage is $1$ minus the square of both lineages disappearing. The second term describes the probability of migrating one lattice site to the left or right, respectively. The first term describes the case of nothing happening in the time interval $\epsilon$.

Letting $\epsilon\to 0$ and performing the Taylor expansion in $\delta x$, we obtain eq.~5 in the main text
\begin{equation}
\partial_t u=D\partial_x^2 u + v\partial_x u +s(x)u-a(x) u^2,
\end{equation}
where $s(x)=a(x)-b(x)$, $D=(v_++v_-)\delta x/2$, and $v=(v_+-v_-)\delta x$. Note that the sign of the convection term $v \partial_x u$ is reversed from that in the equation governing the population density, eq.~5 in the main text.

\subsection{Step-like concentration profile without flow}
In a step-like gradient without flow, the steady-state wild=type population density profile $c(\xi)$ obeys the equation
\begin{equation}
0=\partial_\xi^2 c+ (1-2\Theta(\xi))c - c^2,
\end{equation}
where $\Theta(\xi)$ is the Heaviside function. For both $\xi>0$ and $\xi<0$, this equation can be solved by a mechanical analogy with a particle in the "potential" $U(c)=\pm\frac{c^2}{2} - \frac{c^3}{3}$. For $\xi<0$, we have the potential $U(c)=c^2/2 - c^3/3$ and the boundary condition $c(-\infty)=1$ and $c(0)=c_0$, which gives a total energy $E=K+U=1/6$ because the kinetic energy $K=0$  at $-\infty$. The population density $c(\xi)$ is then determined through the integral
\begin{equation}
\xi=\int^{c(\xi)}_{c_0} \frac{dc'}{\sqrt{-2U(c')+2E}}=\int^{c(\xi)}_{c_0} \frac{dc'}{\sqrt{-c^2+c^3/3+1/3}}.
\end{equation}
Similarly, for $\xi>0$, we have $U(c)=-c^2/2 - c^3/3$ and $E=0$, and $c(\xi)$ is determined through the integral
\begin{equation}
\xi=\int^{c(\xi)}_{c_0} \frac{dc'}{\sqrt{c^2+c^3/3}}.
\end{equation}
Both integrals can be solved exactly, and the derivatives matched at $\xi=0$. The result is 
\begin{eqnarray}
c(\xi)&=\frac{3}{2}\tanh\left[\frac{\xi-\xi_-}{2}\right]^2-\frac{1}{2}, \quad \xi<0,\\
&=\frac{3}{2} \tanh\left[\frac{\xi+\xi_+}{2}\right]^2-\frac{3}{2}, \quad \xi\geq 0,
\end{eqnarray}
where $\xi_\pm=2\mbox{arctanh}(\frac{1}{3}\sqrt{6\pm 3 +\sqrt{6}})$. 

The population density transitions from 1 to 0 exponentially fast, and we can approximate $c(\xi)\approx \Theta(-\xi)$ when computing the establishment probability, which then approximately obeys the equation
\begin{equation}
0=\partial_\xi^2 u+ \Theta(\xi) u - u^2.
\end{equation}
Using the same mechanical analogy as above, and again matching derivatives at $\xi=0$, we find
\begin{eqnarray}
u(\xi)&=\frac{1/\sqrt{3}}{\left(1-\xi/\sqrt{6\sqrt{3}}\right)^2}, \quad \xi<0,\\
&=\frac{3}{2} \tanh\left[\frac{\xi+\xi_u}{2}\right]^2-\frac{1}{2}, \quad \xi\geq 0,
\end{eqnarray}
where $\xi_u=2\mbox{arctanh}(\frac{1}{3}\sqrt{3 +2\sqrt{6}})$.

From the profiles for $c(\xi)$, the total death rate $B$ can directly be computed as $B=-3 + \sqrt{9 + \sqrt{6}}\approx 0.38$, as quoted in the main text. Since $c(\xi)$ transition sharply around $\xi=0$, the establishment score $R=\int_{-\infty}^\infty c(\xi)u(\xi)$ can be approximated as $R\approx \int_{-\infty}^0 u(\xi)=\sqrt{2\sqrt{3}}\approx 1.86$, which is very close to the numerical result $R\approx 1.91$.

\begin{figure}
\centering\includegraphics[width=0.6\textwidth]{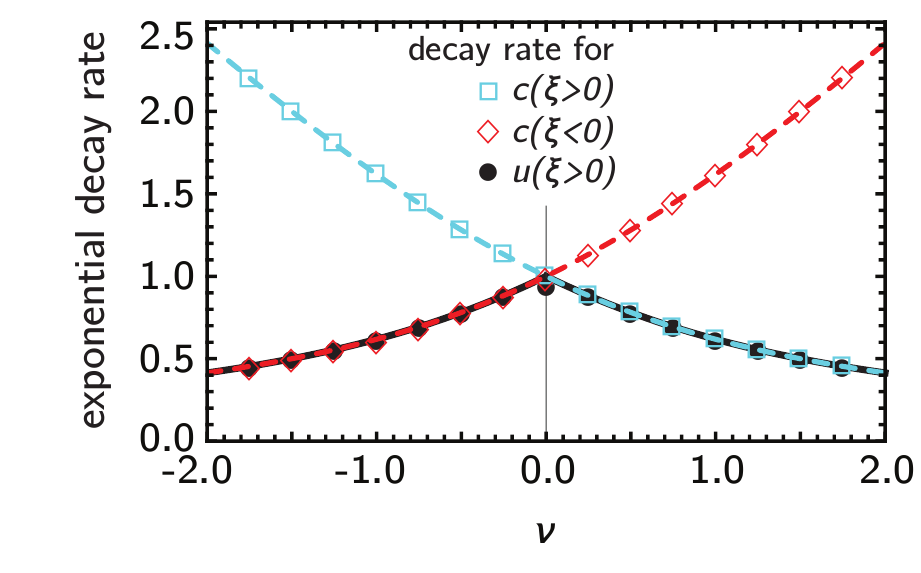}
\caption{Exponential decay rates for the different cases discussed in the text: $c(\xi>0)$ (cyan, eq.~\ref{eq:c}), $c(\xi<0)$ (red, eq.~\ref{eq:oneminusc}), and $u(\xi>0)$ (black, eq.~\ref{eq:u}).}
\label{fig:exponents}
\end{figure}

\subsection{Step-like antibiotic concentration profile with flow }
In the presence of flow, the equations for the population density becomes
\begin{equation}
0=\partial_\xi^2 c - \nu \partial_\xi c + (1-2\Theta(\xi))c - c^2.
\end{equation}
For $\xi \gg 0$, the population density is small such that we can extract the asymptotic behavior from the linear equation. With the boundary condition $c(\infty)=0$, we get
\begin{equation}
c(\xi\gg 0) \sim e^{-(\sqrt{4+\nu^2}-\nu)\xi/2}.
\label{eq:c}
\end{equation}
Similarly, for $\xi \ll 0$, $c(\xi)$ is close to one; introducing $\Psi(\xi)=1-c(\xi)$ and linearizing the result equation, we get that
\begin{equation}
1- c(\xi \ll 0) \sim e^{-(\sqrt{4+\nu^2}+\nu)|\xi|/2}.
\label{eq:oneminusc}
\end{equation}
Thus, even in the presence of flow, the population density transitions sharply around the antibiotic concentration step. Therefore, we can employ the same approximation as in the no-flow case and approximate $c(\xi)\approx \Theta(-\xi)$. 

The establishment probability $u(\xi)$ thus approximately satisfies the equation
\begin{equation}
0=\partial_x^2 u+\nu \partial_x u + \Theta(\xi) u - u^2.
\label{eq:uwithflow}
\end{equation}
For $\xi>0$, the establishment probability approaches 1 exponentially quickly. To see this, we can again expand around 1 and solve the resulting linearized equation. The result is 
\begin{eqnarray}
1- u(\xi>0) &\sim e^{-(\sqrt{4 + \nu^2}-\nu) \xi/2}, \quad\mbox{for }\nu > 0,\\
 &\sim e^{-(\sqrt{4 + \nu^2}+\nu) \xi/2}, \quad\mbox{for }\nu < 0.
\label{eq:u}
\end{eqnarray}
The results in summarized in Fig.~\ref{fig:exponents}. 

For $\xi<0$, the linear term in eq.~\ref{eq:uwithflow} vanishes, and we have to distinguish two cases, $\nu>0$ and $\nu<0$. For $\nu>0$, we expect a broad profile because co-flow facilitates the establishment of mutations born deep inside the population bulk. Hence, the non-linear term in eq.~\ref{eq:uwithflow} cannot be neglected. On the other hand, if the co-flow is strong, we can neglect the diffusive term and find asymptotically
\begin{equation}
u(\xi<0) \sim \frac{1}{1+\frac{|\xi|}{\nu}}, \quad\mbox{for }\nu > 0,
\end{equation}
in good agreement with the numerical solution (cyan line in Fig.~4b in the main text). 

For counter-flow, i.e., for $\nu<0$, it is intuitively clear that the establishment probability must vanish faster than in the no-flow case. Thus, we expect for that $u(\xi \ll 0)$ such that we can neglect the non-linearity. As a result, we get 
\begin{equation}
u(\xi<0) \sim e^{-\nu \xi}, \nu <0,
\end{equation}
which also agrees well with the numerical solution (red line in Fig.~4b in the main text).

\begin{figure}
\centering\includegraphics[width=0.8\textwidth]{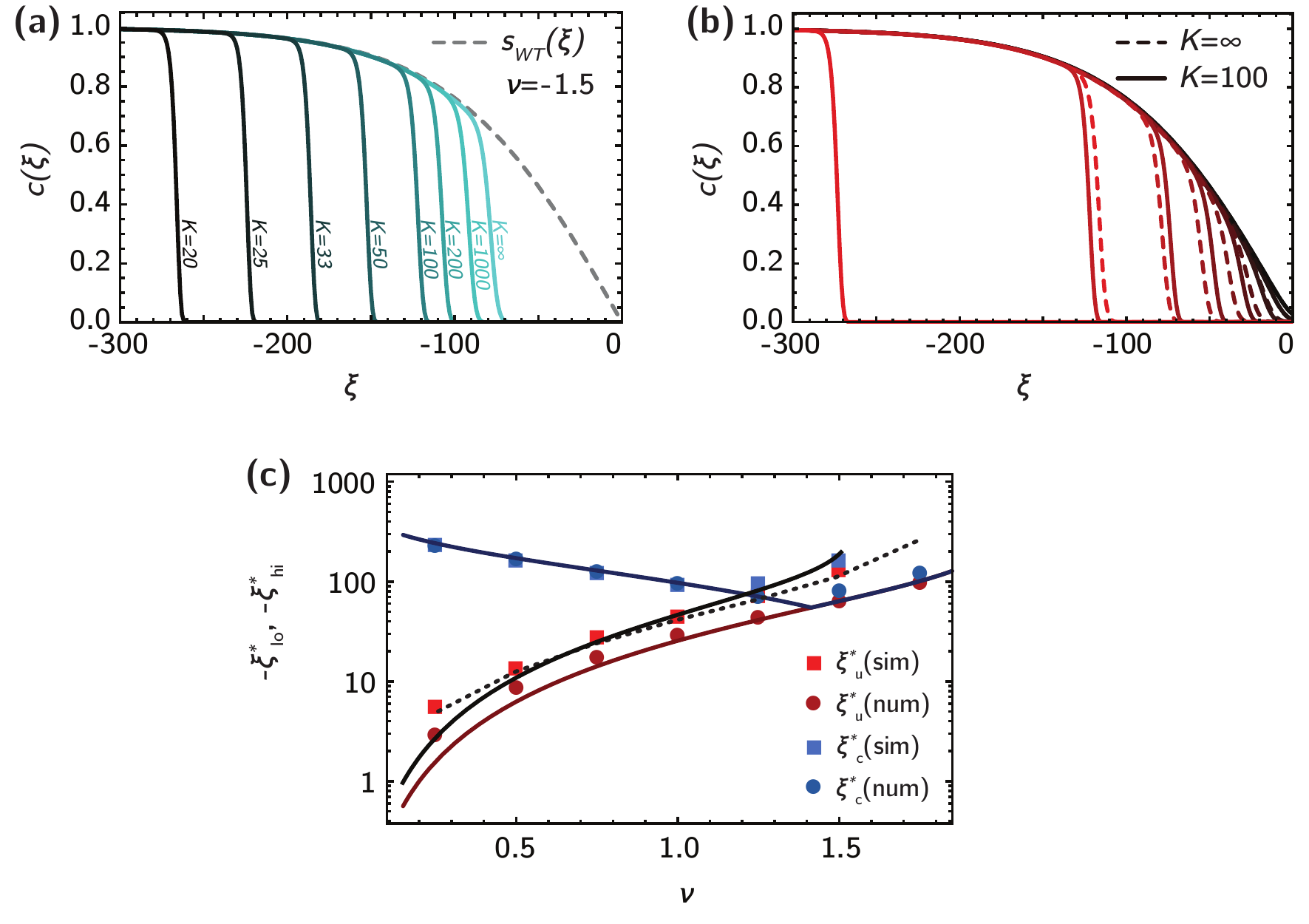}
\caption{The effects of finite carrying capacity on the population density profile in shallow gradients. (a) In counter-flow conditions ($\nu=-1.5$), a small carrying capacity leads to a strong shift in the cut-off. (b) Evaluation the analytical model with the growth rate corrected for finite carrying capacity, eq.~\ref{eq:finiteK} (solid lines), and without the correction (dashed lines) for different flow speeds. (c) Comparison with the simulation shows good agreement between the numerical data (dashed line) and the analytical approximation, eq.~\ref{eq:xicK} (solid line).}
\label{fig:Keffect}
\end{figure}

\subsection{Cut-off position and finite carrying capacity}
Here, we provide a more rigorous derivation of the cut-off position $\xi_c^*$. For $\nu <0 $ and $m \ll 1$, the population density $c(\xi,\tau)$ satisfies the equation 
\begin{equation}
\partial_\tau c=\partial_\xi^2 c - \nu \partial_\xi c(\xi) +  s_{WT}(\xi) c -c^2.
\end{equation}
Performing the transformation $\varphi=ce^{-\nu\xi/2}$ we obtain
\begin{equation}
\partial_\tau \varphi(\xi,\tau) = \partial_{\xi}^2 \varphi + \left(s_{WT}(\xi)-\frac{\nu^2}{4}\right)\varphi- \varphi^2 e^{-\nu\xi/2}.\label{eq:fulltimedepsigma}
\end{equation}
Assume the population is started with an initial population at $\xi=-\infty$. To arrive at the steady-state population density discussed in the main text, $\partial_\tau \varphi$ must be positive initially and go to zero as $\tau \to \infty$. For the right-hand side of eq.~\ref{eq:fulltimedepsigma} to be positive, we must have $s_{WT}(\xi)-\frac{\nu^2}{4}>0$, whence we obtain eq.~15 in the main text. Eq.~17 is obtained by the same arguments. Note that this derivation also gives the general condition $\nu^2<4$, i.e., the external flow speed must be smaller than the Fisher wave speed of the population.

We model the effects of a finite carrying capacity $K$ by introducing a cut-off in the net growth rate if the local wild-type population density becomes lower than $1/K$. To this end, we replace $s_{WT}(\xi)$ in eq.~5 in the main text with
\begin{equation}
s_{WT}(\xi, K)=s_{WT}(\xi)\Theta(c(\xi)-1/K).
\label{eq:finiteK}
\end{equation}
The results are shown in Fig.~\ref{fig:Keffect}: for fixed $\nu$, smaller $K$ reduce the maximal sustainable local flow speed, such that the cut-off is shifted to the left for smaller $K$ (Fig.~\ref{fig:Keffect}a). When $K$ is relatively small, the cut-offs can be shifted strongly, especially at strong flow speeds (Fig.~\ref{fig:Keffect}b). 

For an analytical approximation, we use the result that a finite carrying capacity reduced the Fisher wave speed by a factor $1-\pi^2/2(\ln K)^2$. Plugging this correction into eq.~16 we get a corrected cut-off position $\xi^*_{c}$,
\begin{equation}
\xi^*_{c}(K)=-\frac{1}{m}\rm{arctanh}\left[\frac{\nu^2}{4\left(1-\pi^2/2(\ln K)^2\right)^2}\right],
\label{eq:xicK}
\end{equation}
which shows reasonable agreement with our simulations (Fig.~\ref{fig:Keffect}c, black line).

\subsection{Small mutation rate estimate}
For clonal interference to be negligible, we estimate that the time to fixation of a mutant clone must be much faster than the time it takes for a mutation to arise and establish. In a well-mixed population, this condition can be phrased as $\mu N \ll 1$, or, more precisely, 
\begin{equation}
\tau_{\mbox{\tiny{fix}}}\approx \ln[Ns]/s \ll \tau_{\mbox{\tiny{establish}}} \approx 1/(\mu N s). 
\end{equation}

In a spatial scenario, the establishment time $\tau_{\mbox{\tiny{establish}}}$ is given through a generalization of the well-mixed result, i.e.,
\begin{equation}
\tau_{\mbox{\tiny{establish}}}^{-1} \approx \mu c_\infty \! \int u(x)c(x) dx=\mu (K/\ell) R,
\end{equation}
where we have made the dependence on the carrying capacity $K$ explicit, and the mutation rate $\mu$ is given per generation. We have found in the main text that $R\sim \ell$ in steep gradients and $R \sim \lambda$ in shallow gradients.

We can estimate the time to fixation by considering a mutation born at a typical distance $L$ from the front of the population; it then has to travel this distance $L$ to take over the whole population. In our model, the distance $L$ can be estimated from the position of the peak of the product $u(\xi)c(\xi)$. This is approximately given by
\begin{eqnarray}
L\sim &\ell, \quad \lambda \ll \ell,\\
&\lambda, \quad \lambda \gg \ell. 
\end{eqnarray}
The time it takes to travel this distance depends on the steepness of the gradient: in a steep gradient, mutants have to migrate into region of high antibiotic concentration through random dispersal. We can model their motion as a random walk, and hence, the distance $\ell = \sqrt{2 D \tau_{\mbox{\tiny{fix}}}}$ such that $\tau_{\mbox{\tiny{fix}}}=\ell^2/2D$. This gives the criterion for negligible clonal interference as 
\begin{equation}
\mu K \ll 1.
\end{equation}

In shallow gradients, the time to fix for a mutant clone is given by the length $L$ divided by the speed $v$ with which it travels, which is roughly $v\sim \sqrt{D a_0}$. This gives the condition 
\begin{equation}
\mu K \ll \frac{\ell^2}{\lambda^2}.
\end{equation}

\bibliographystyle{iopart-num}

\providecommand{\newblock}{}

\end{document}